\documentclass[amsmath,amssymb,graphicx,11pt]{article} 

\usepackage{amsmath, amssymb, graphicx}

\usepackage[english]{babel}
\usepackage[autostyle, english = american]{csquotes}
\MakeOuterQuote{"}

\usepackage{chngcntr}
\counterwithin{equation}{section}

\newcommand{\finbox}{\hspace*{\fill}$\rule{0.17cm}{0.17cm}$}
\newcommand{\FBOX}{\hspace*{\fill}$\rule{0.17cm}{0.17cm}$}
\newcommand{\BB}{\hspace*{\fill}$\rule{0.17cm}{0.17cm}\
\rule{0.17cm}{0.17cm}$}

\newcommand{\finboxHere}{\ $\rule{0.17cm}{0.17cm}$}
\newcommand{\BOX}{\ $\rule{0.17cm}{0.17cm}$}

\def\eref#1{(\ref{#1})}

\def\emeref#1{ {\em(\ref{#1})} } \def\emeref#1{ {\em(\ref{#1})}}

\def\emref#1{ {\em\ref{#1}} } \def\emref#1{ {\em\ref{#1}}}

\newcommand{\Proof}{\noindent {\bf Proof.  }}

\newtheorem{THM}{THEOREM}[section] \newtheorem{LEMMA}[THM]{Lemma}
\newtheorem{CLAIM}[THM]{Claim} \newtheorem{COR}[THM]{Corollary}
\newtheorem{PROP}[THM]{Proposition}

\newtheorem{Thm}[THM]{Theorem} \newcommand{\Theorem}{\begin{Thm}}
\newcommand{\eTh}{\end{Thm}}

\newtheorem{CON}[THM]{Conjecture}
\newcommand{\Conjecture}{\begin{CON}} \newcommand{\eCon}{\end{CON}}

\newtheorem{CONS}[THM]{} \newcommand{\Conjectures}{\begin{CONS}}
\newcommand{\eCons}{\end{CONS}}

\newcommand{\eq}{\begin{equation}} \newcommand{\eeq}{\end{equation}}

\newcommand{\THEOREM}{\begin{THM}} \newcommand{\eT}{\end{THM}}
\newcommand{\Lemma}{\begin{LEMMA}} \newcommand{\eL}{\end{LEMMA}}
\newcommand{\Claim}{\begin{CLAIM}} \newcommand{\eCl}{\end{CLAIM}}
\newcommand{\Corollary}{\begin{COR}} \newcommand{\eCo}{\end{COR}}

\newcommand{\Proposition}{\begin{PROP}} \newcommand{\eP}{\end{PROP}}

\newtheorem{EXAM}{Example}[section]

\newcommand{\Example}{\begin{EXAM}} \newcommand{\eXa}{\end{EXAM}}

\newtheorem{EX}{Exercise}[section]

\newcommand{\Exercise}{\begin{EX}} \newcommand{\eEx}{\end{EX}}
\newtheorem{EXS}[EX]{} \newcommand{\Exercises}{\begin{EXS}}
\newcommand{\eExs}{\end{EXS}}

\newtheorem{PROBLEM}[EX]{Problem}
\newcommand{\Problem}{\begin{PROBLEM}}
\newcommand{\ePm}{\end{PROBLEM}}

\newtheorem{PROB}[EX]{} \newcommand{\Prob}{\begin{PROB}}
\newcommand{\eProb}{\end{PROB}}

\newtheorem{PROBLEMS}[EX]{} \newcommand{\Problems}{\begin{PROBLEMS}}
\newcommand{\ePms}{\end{PROBLEMS}}

\newtheorem{OPROBLEM}[EX]{Research problem}
\newcommand{\Open}{\begin{OPROBLEM}} 
\newcommand{\eO}{\end{OPROBLEM}}

\newtheorem{REMARK}{Remark}[section] 

\newcommand{\Remark}{\begin{REMARK}} 
\newcommand{\eRe}{\end{REMARK}}

\newtheorem{COMMENT}{Comment}[section]
\newcommand{\Comment}{\begin{COMMENT}} 
\newcommand{\eCom}{\end{COMMENT}}

\newtheorem{TODO}{Todo}[section] \newcommand{\Todo}{\begin{TODO}}
\newcommand{\eTo}{\end{TODO}}

\newtheorem{MEMO}{Memo}[section] \newcommand{\Memo}{\begin{MEMO}}
\newcommand{\eMe}{\end{MEMO}}

\newtheorem{ALGORITHM}[THM]{Algorithm}
\newcommand{\Algorithm}{\begin{ALGORITHM}}
\newcommand{\eAl}{\end{ALGORITHM}}

\begin{document}


\title {How to see the forest despite the trees}

\author{Erika B\'erczi-Kov\'acs\thanks{Department of Operations
Research, ELTE E\"otv\"os Lor\'and University, P\'azm\'any P. s. 1/c, Budapest, Hungary, and
HUN-REN-ELTE Egerv\'ary Research Group e-mail:  {\tt
erika.berczi-kovacs\char'100 ttk.elte.hu } } \ and \ {Andr\'as
Frank\thanks{ Department of Operations
Research, ELTE E\"otv\"os Lor\'and University, P\'azm\'any P. s. 1/c, Budapest, Hungary, and
HUN-REN-ELTE Egerv\'ary Research Group, e-mail:
{\tt andras.frank\char'100 ttk.elte.hu } } } }

\date{July 2, 2026}

\medskip

\maketitle

\begin{abstract} One of the major starting points of discrete
optimization is the theorem of Nash-Williams and Tutte on the
existence of $k$ disjoint spanning trees of a graph, along with its
counterpart on the existence of $k$ forests covering all edges of the
graph.  These elegant results triggered comprehensive research that
gave rise to far-reaching generalizations and found applications in
seemingly distant areas.  

Our first goal is to elucidate some aspects of these
developments with the hope that the story finds its way to
non-experts.  But we hope that experts will also find some novelty
in our exposition.  \end{abstract}

\medskip \medskip


\section{Introduction}

Graphs (directed, undirected, mixed, hyper-) are fundamental
mathematical tools for modelling problems of real life networks.  For
example, a GPS seeks to find the fastest/shortest path from one
location on a map to another.

For managing practical problems, the very first step is to capture
(define) formally certain intuitive features of the network.  For
example, the intuition that a graph $G=(V,E)$ with node-set $V$ and
edge set $E$ \lq consists of one piece/component\rq\ or that $G$ \lq
is connected\rq\ can be formally defined by requiring that $V$ cannot
be cut (partitioned) into two non-empty parts in such a way that no
edge connects the two parts.  Another possibility is to require that
there is a path connecting any two nodes.  These two formal
definitions of connectedness are easily seen to be equivalent.

A {\bf tree} $G=(V,T)$ is a critically (or minimally) connected graph
in the sense that deleting any of its edges destroys connectivity.
This is equivalent to requiring that $G$ is connected and includes no
circuit.  Yet another equivalent definition is that $G$ is connected
and every non-empty subset $X\subseteq V$ of nodes is sparse in the
sense that $X$ induces at most $\vert X\vert -1$ edges.  The
connectedness of a graph $G$ is equivalent to requiring that $G$
includes a spanning tree.  A {\bf forest} is a graph whose connected components
are trees.  An equivalent definition is that the graph includes no
circuit, that is, the graph is sparse. In what follows, sometimes we refer to the edge-set itself of a forest as a forest.

Node, edge, degree, path, circuit, loop, cut, tree, forest,
arborescence, branching, (connected) component:  these are some of the
most important basic concepts of graph theory.  An easy property is
that an undirected graph $G=(V,E)$ with an $n$-element node-set $V$
and edge set $E$ admits an $st$-path (a path connecting two specified
nodes $s,t\in V$) if and only if there is no empty cut separating $s$
and $t$, that is, the number $d_G(X)$ of edges between $X$ and $V-X$
is positive whenever $s\in X\subseteq V-t$.  There are simple
efficient algorithms for checking local connectivity (between nodes
$s$ and $t$) that 
either find such a set $X$ with $d_G(X)=0$ or find an $st$-path.  $G$ is called (globally) connected if there is a
path between any two nodes $s,t\in V$.

A basic graph optimization problem is that of finding a spanning tree of $G$
which is cheapest with respect to a cost function $c: E\to \mathbb{R}$.  A simple
greedy algorithm, due to Kruskal, does this job~\cite{Kruskal}.  By Dijkstra's
classic algorithm~\cite{Dijkstra}, the problem of finding algorithmically a cheapest
$st$-path is also efficiently manageable, at least for non-negative
cost functions.  However, the sensitivity of this class of problems is
nicely reflected by the fact that the cheapest (or shortest) path
problem is NP-hard for general cost-functions~\cite{GareyJohnson}.  On the positive side,
when negative cost values are allowed and only circuits with negative
total costs are forbidden, then there is a polynomial algorithm, which
relies, however, on deeper ideas.

The need for capturing intuitively more \lq massive/strong/solid\rq \
connectivity concepts appeared as early as 1927, when Menger
characterized those graphs admitting $k$ internally disjoint $st$-paths \cite{Menger}.
Here it is straightforward to figure out a natural necessary
condition:  no $k-1$ nodes (distinct from $s$ and $t$) can block all
$st$-paths (we can assume $s$ and $t$ are not adjacent).  The deep part of Menger's theorem is the proof of the
sufficiency of this condition.

Naturally, there are several other variants for extending connectivity
concepts.  For example, a variation of Menger's theorem concerns the
existence of $k$ edge-disjoint $st$-paths, which is often referred to
as the (undirected) edge-Menger theorem.  These results gave rise to a
huge discipline:  network flows (in directed graphs), in particular,
the max-flow min-cut theorem.  It should be emphasized that in this
area it is rather typical that figuring out a necessary condition for
the existence of the investigated property is easy, and the essence is
finding a (possibly algorithmic) proof of sufficiency (see, for
example, Hall's theorem on the existence of a perfect matching in a
bipartite graph, or Dilworth's theorem on the partionability of a
poset into $k$ chains~\cite{Dilworth}).

Global connectivity has also been extended to higher connectivities.
A graph $G$ is {\bf $k$-edge-connected} if $d_G(X)\geq k$ whenever
$\emptyset \subset X\subset V$, which is equivalent, by the edge-Menger
theorem, to requiring that there are $k$ edge-disjoint $st$-paths for
every pair $\{s,t\}$ of nodes.  $G$ is {\bf $k$-node-connected} if it
has at least $k+1$ nodes and $G$ remains connected after leaving out
any subset of less than $k$ nodes.

In this line of possible extensions of connectivity concepts, a natural
question arises:  what would be a correct generalization of the above
mentioned trivial statement that a graph is connected if and only if
it has a spanning tree?

A graph is called {\bf $k$-tree-connected} if it contains $k$
edge-disjoint spanning trees.  Following the theme of the edge-Menger
theorem, one might surmise at first sight that a graph is
$k$-edge-connected if and only if it is $k$-tree-connected.  However,
this is plainly wrong even for $k=2$, as is shown by a triangle.
This toy example indicates that one must also require that the graph
should have at least $k\cdot (\vert V\vert -1)$ edges.  However, even
this strengthened necessary condition is not sufficient, and the
question arises:  what is a necessary and sufficient condition for a
graph to be $k$-tree-connected?

For Menger (also for Hall \cite{Hall-thm} or Dilworth), it was easy to formulate a
necessary condition. For the problem of finding $k$ disjoint spanning
trees, however, just figuring out a promising conjecture is already a
great challenge.  A bit surprisingly, the answer was provided by two
authors, W.T.  Tutte and C.St.J.A Nash-Williams, independently of each
other.  Their papers appeared in the same issue (number 36) of the
Journal of the London Mathematical Society, in 1961.  The paper of Tutte
was submitted on April 29, 1960, and the paper of of Nash-Williams was
submitted on December 4, 1960.

Tutte writes on the third page of his paper:  \medskip

"Note.--Dr.\ C.\ St.\ J.\ A.\ Nash-Williams has sent me another proof of
Theorem I (obtained quite independently), which does not use the
concept of a multiple graph."

\medskip

Nash-Williams writes at the beginning of his paper:

\medskip

"Tutte [3] proved a theorem equivalent to Theorem 1 of the present
paper.  A few months later, the present author (unaware of Tutte's
work) obtained a different proof, which is the object of this paper to
present."

\medskip

Using notation fitting to the present overview, Theorem I of Tutte
and Theorem 1 of Nash-Williams are as follows.

\THEOREM [Tutte \cite{Tutte61a}, {\rm Theorem I}] \label{Tutte-pakol1}
An undirected graph $G=(V,E)$ includes $k$ edge-disjoint spanning
trees if and only if, for every subset $F\subseteq E$ of edges, 

\eq \vert F\vert \ \geq \ k \cdot (q(G,F)-1), \label{(Tutte-pakol1)}
\eeq

\noindent where
$q(G,F)$ denotes the number of connected components of the subgraph of
$G$ induced by $E-F$.  \eT

We hasten to emphasize that Tutte actually proved this result in a
significantly more general form (Theorem II), see Theorem
\ref{Tutte-pakol2} below.

\THEOREM [Nash-Williams \cite{Nash61}, {\rm Theorem 1}]
\label{Nash-pakol} A graph $G$ has $k$ edge-disjoint spanning trees if
and only if

\eq e_G({\cal P}) \geq k\cdot (\vert {\cal P}\vert -1)
\label{(Nash-pakol)} \eeq

\noindent for every partition $\cal P$ of $V(G)$, where $e_G({\cal
P})$ denotes the number of edges connecting distinct members of $\cal
P$.  \eT

We assume throughout that graphs are loopless.  It is also
assumed that a partition consists of non-empty sets and has at least
two members.  In what follows, we shall refer to the edges connecting
distinct parts of a partition $\cal P$ as {\bf cross-edges} of $\cal
P$.

In an equivalent formulation, the condition in Theorem \ref{Nash-pakol} is that the
number of cross-edges of every $q$-partite ($q=2,3,\dots ,n$)
partition of $V(G)$ is at least $k\cdot (q-1)$.  Such graphs are
sometimes called {\bf $k$-partition-connected}.

It is seen immediately, without referring to the conclusions of the two theorems,
that the conditions in the theorems are equivalent.  Namely, the
Tutte condition may be viewed as a redundant form of the Nash-Williams
condition.  In fact, even the latter is somewhat redundant since it
suffices to require \eref{(Nash-pakol)} only for partitions where each
part induces a $(k+1)$-edge-connected subgraph.

The form of Theorem \ref{Nash-pakol} immediately implies that a
$2k$-edge-connected graph $G=(V,E)$ is $k$-tree-connected and moreover remains so after deleting an arbitrary subset $K$ of at most $k$
edges.  Indeed, for a partition $\{V_1,\dots ,V_q\}$ of $V$, the
number of cross-edges in $G':=(V,E-K)$ is $\sum _id_{G'}(V_i)/2 \geq
\sum _id_{G}(V_i)/2 -k = q(2k)/2 - k =k\cdot (q-1)$, and Theorem
\ref{Nash-pakol} can be applied to $G'$ in place of $G$.

The theorems of Tutte and Nash-Williams can be formulated in a concise
way, as follows.

\THEOREM [Tree-packing theorem of Tutte and Nash-Williams] \label{TNW}
A graph $G$ is $k$-tree-connected if and only if $G$ is
$k$-partition-connected.\eT

For later generalization, it is useful to introduce a natural
extension of $k$-partition-connecti\-vi\-ty \cite{FrankJ49}.  Beside the
non-negative integer $k$, let $l$ be a non-negative integer.  A graph
$G=(V,E)$ is called {\bf $(k,l)$-partition-connected} if, for any partition $\cal P$ of $V$,

\eq e_G({\cal P}) \geq k\cdot (\vert {\cal P}\vert -1) + l
\label{(Nash-pakol2x)}. \eeq

\noindent  Obviously,
$(k,0)$-partition-connectivity is the same as
$k$-partition-connectivity, $(0,l)$-partition-connectivity is
equivalent to $l$-edge-connectivity, and
$(k,k)$-partition-connectivity is $2k$-edge-connectivity, or more
generally, for $k\leq l$, $(k,l)$-partition-connectivity is equivalent
to $k+l$-edge-connectivity.  Theorem \ref{TNW} immediately shows that for a
given non-negative integer $l$, a graph $G$ remains
$k$-tree-connected after leaving out any subset of $l$ edges if and
only if $G$ is $(k,l)$-partition-connected.

This observation on $2k$-edge-connected graphs has far-reaching
applications. For example, it served as a starting point for the
design of min cut algorithms \cite{Karger2000} and remained the basis
for many others.  Karger \cite{Karger2000} observed that given a
maximum set of edge-disjoint spanning trees, a minimum cut intersects
at least one of them in at most 2 edges. Thus to find a minimum cut it
is enough to find the minimum of those cuts containing at most 2 edges
from one of the given trees.

A natural counterpart of the disjoint tree theorem investigates the
problem of decomposing the edge set of a graph $G=(V,E)$ into $k$
forests, which is equivalent to covering $E$ by $k$ spanning trees when
$G$ is connected.  This was solved by Nash-Williams in a (one page) paper that also appeared in the Journal of the London Mathematical 
Society.

\THEOREM [Tree-covering theorem of Nash-Williams \cite{Nash64}]
\label{Nash-fed} The edge set of a connected graph $G=(V,E)$
can be covered by $k$ spanning trees (or equivalently, $E$ can be
partitioned into $k$ forests) if and only if $G$ is {\bf
$k$-forest-sparse} (or just $k$-sparse) in the sense that

\eq i_G(X) \leq k\cdot (\vert X\vert -1) \label{(k-sparse)} \eeq

\noindent for every non-empty subset $X$ of $V$, where $i_G(X)$
denotes the number of edges induced by $X$.  \eT

\noindent {\bf Proof outline} \ The necessity is straightforward,
while sufficiency follows once we observe that $G$ may be assumed to
be saturated in the sense that adding any new edge destroys
$k$-sparsity.  For such a graph the supermodularity of $i_G$ (i.e. for every pair of subsets $X,Y: i_G(X)+i_G(Y)\leq i_G(X\cap Y)+i_G(X\cup Y)$) implies
that $\vert E\vert = k\cdot (\vert V\vert -1)$, in which case
$k$-sparsity is just equivalent to $k$-partition-connectivity, and
then the tree-packing theorem implies the tree-covering theorem.
\FBOX

\medskip

For integers $0\leq l<k$, we call a graph $G=(V,E)$ {\bf
$(k,l)$-forest-sparse} if the graph obtained from $G$ by adding any
set of $l$ new edges is decomposable into $k$ forests.  By Theorem
\ref{Nash-fed} of Nash-Williams, this is equivalent to requiring that, for every subset $X$ of $V$ with $\vert X\vert \geq
2$,

\eq i_G(X) \leq k\cdot (\vert X\vert -1) - l \label{(kl-sparse)}. \eeq

\noindent   In the literature (e.g.  \cite{Lee+Streinu2008}) a graph is
called $(k,l)$-sparse if, for
every nonempty subset $X\subseteq V$ (where $0\leq l\leq 2k-1$), $i_G(X)\leq k\cdot\vert X\vert - l$.
Clearly, a graph is $(k,l)$-forest-sparse if and only if it is
$(k,k+l)$-sparse.

It should be noted that there is yet another closely related paper
that appeared in the Journal of London Mathematical  Society, actually
earlier than the above-mentioned works of Nash-Williams and Tutte.  A.
Horn proved the following.

\THEOREM [Horn \cite{Horn55}] \label{Horn} A (finite) set $S$ of
vectors can be partitioned into $k$ linearly independent subsets if 
and only if, for every subset $X$ of $S$, \eq k\cdot r(X)\geq \vert X\vert, \label{(Horn)} \eeq
 where $r(X)$ denotes the rank of $X$.
\eT

Somewhat surprisingly, the same theorem was published in the same journal
by R. Rado in 1962 \cite{Rado1962a}, without referring to Horn's
paper.  (For an in depth historical overview, see section 42.6f of
Schrijver's book  \cite{Schrijverbook}).

It is also known for a graph $G=(V,E)$
that there is a matrix $M_G$ whose columns correspond to the edges of
$G$ in such a way that a subset $X\subseteq E$ is a forest if and only
if the set of columns of $M_G$ corresponding to $X$ is linearly
independent.  These two observations show that Horn's theorem implies
Theorem \ref{Nash-fed}.

\medskip

These initial results of Nash-Williams and Tutte on trees and forests
triggered immense advancements in various related areas:  matroid
optimization, directed counterparts, hypergraph extensions.  A rich
overview of the area can be found, for example, in the books of Frank
\cite{Frank-book} and Schrijver \cite{Schrijverbook}, and there are
thousands of papers referring to the above-mentioned results of
Nash-Williams and Tutte.  
Obviously, the present summary is not a suitable venue in which to provide a
general overview, and our goal is only to  outline some of
the most exciting results.

\subsection{Some graph problems triggered by the tree-packing and
tree-covering theorems} \label{Problems}

We close this introductory section by a list of naturally emerging
questions related to the tree-packing and the tree-covering theorems,
which will be answered in the remaining sections of this paper.  Some notions used
here are defined later.

\Problem \label{algo1} {\em How can we find (algorithmically) $k$
disjoint spanning trees, when they exist?  How can we find a deficient
partition (that is, one violating \eref{(Nash-pakol)}) when no $k$
disjoint spanning trees exist?  In the latter case, what is the
minimum number (total cost) of new edges whose addition makes the
graph $k$-tree-connected? } \ePm

\Problem \label{extend} {\em Given $k$ edge-disjoint forests
$F_1,\dots ,F_k$ of a graph $G=(V,E)$, when is it possible to extend
these forests \ (A) \ to $k$ edge-disjoint spanning trees, \ (B) \ to
$k$ forests covering all edges?  More generally, for given edge sets
$E_1,\dots ,E_k$ with $F_i\subseteq E_i$, one may require that the
extension of $F_i$ be a subset of $E_i$ ($i=1,\dots ,k$).  
} \ePm

\Problem \label{bounds} {\em Given non-negative lower bounds
$f_1,\dots ,f_k$ and upper bounds $g_1,\dots ,g_k$, when do there
exist $k$ disjoint forests $F_1,\dots ,F_k$ in $G$ for which $f_i\leq
\vert F_i\vert \leq g_i$ for $i=1,\dots ,k$?  } \ePm

\Problem \label{common1} {\em What is the maximum cardinality $K$ of
the union of $k$ forests of $G$?  Clearly, $K=\vert E\vert $ precisely
when $G$ can be decomposed into $k$ forests, and $K=k\cdot (\vert
V\vert -1)$ when $G$ includes $k$ spanning trees.  In an equivalent
formulation, what is the minimum number of new edges which are
parallel to existing ones, whose addition to $G$ results in a
$k$-tree-connected graph?  } \ePm

\Problem \label{weighted} {\em Given a non-negative weight-function on
$E$, what is the maximum total weight of the union of $k$ forests?
Given a non-negative cost-function on $E$, what is the minimum total
cost of the union of $k$ disjoint spanning trees?  } \ePm

\Problem \label{multi-cost} {\em Given $k$ cost-functions $c_1,\dots
,c_k$ on $E$, find $k$ disjoint spanning trees $F_1,\dots ,F_k$
for which the total cost $\widetilde c_1(F_1)+ \widetilde
c_2(F_2)+\dots +\widetilde c_k(F_k)$ is minimum, where $\widetilde
c_i(F_i):= \sum [c_i(e):e\in F_i]$.   } \ePm

\Problem \label{hypergraph} {\em What are the counterparts of the
theorems of Nash-Williams and Tutte concerning hypergraphs?   } \ePm

\Problem \label{digraph1} {\em What are the counterparts of the
theorems of Nash-Williams and Tutte concerning directed and mixed
graphs?  } \ePm

\Problem \label{min-cost-digraph1} {\em (A directed counterpart of the
second part of Problem \ref{weighted}) For a given cost-function, how
can we compute a cheapest rooted $k$-arc-connected subgraph of a
digraph?  } \ePm

\Problem \label{multi-cost-digraph} {\em (A directed counterpart of
Problem \ref{multi-cost}) For given cost-function $c_1,\dots c_k$, how
can we compute $k$ arc-disjoint spanning arborescences $F_1,\dots
,F_k$ with given root of a digraph for which $\widetilde
c_1(F_1)+\cdots +\widetilde c_k(F_k)$ is minimum?  } \ePm

\section{Matroids in the background}

As already indicated, Tutte actually proved the tree-packing theorem
in a more general form.  He called a set of graphs $G_1=(V_1,E,\varphi
_1)$, \dots , $G_k=(V_k,E,\varphi _k)$ a {\bf multiple graph} where
$\varphi _i(e)=\{u,v\}$ denotes the end-nodes of $e\in E$ in $V_i$.

\THEOREM [Tutte \cite{Tutte61a}, {\rm Theorem II.}]
\label{Tutte-pakol2} Suppose that the graphs $G_1=(V_1,E,\varphi _1)$,
\dots , $G_k=(V_k,E,\varphi _k)$ form a multiple graph.  There exist
$k$ disjoint edge sets $F_1\subseteq E,\dots ,F_k\subseteq E$ for
which $F_i$ is a spanning tree of $G_i$ \ $(i=1,\dots ,k)$ if and only
if, for every subset $F\subseteq E$ of edges,

\eq \vert F\vert \ \geq \ \sum _{i=1}\sp k (q(G_i,F)-1),
\label{(Tutte-pakol2)} \eeq

\noindent  where
$q(G_i,F)$ denotes the number of (connected) components of the
subgraph $(V_i,E-F, \varphi _i)$ of $G_i$ induced by $E-F$.  \eT

This theorem rather easily implies, for example, a solution to Part
(A) of Problem \ref{extend}.  A bit surprisingly, despite the numerous
works in the literature citing the paper \cite{Tutte61a} of Tutte,
only very few of them mention Theorem \ref{Tutte-pakol2}.  A reason may be that the notion of Tutte's multiple graph is not
particularly intuitive.

As is typical in mathematics, a phenomenon is more naturally
understandable in an appropriately generalized framework.  In the
present case, the concept of matroids will play this role. What is a matroid?  
A fundamental
property in linear algebra (roughly) is that if we have a (finite) set
$K$ of vectors, then any linearly independent subset of $K$ can be
extended to a largest independent subset of $K$ (whose cardinality is
called the rank of $K$).  A formally similar feature in graph theory
is that a forest of a connected graph can always be extended to a
spanning tree.  The concept of matroid captures this \lq
greedy-type\rq \ property in an abstract way.  Namely, a matroid
$M=(S,{\cal F})$ is a special hypergraph on (a typically finite)
ground-set $S$, where $\cal F$ is a downward closed set system
containing the empty set with the property that, for each $S'\subseteq
S$, the maximal members of $\cal F$ in $S'$ have the same cardinality
(called the rank of $S'$).  The members of $\cal F$ are called \lq
independent\rq \ sets.

Linear independent sets of vectors form a matroid, but it should be
emphasized that not every matroid can be represented by vector spaces.
Another classic example comes from graph theory:  the forests of an
undirected graph $G=(V,E)$ form the independent sets a matroid (on
ground-set $E$).  This is a trivial observation, a much deeper result
tells that the unions of $k$ forests also form a matroid.  Yet another
non-trivial example is the matching matroid:  a subset $Z$ of nodes is
declared independent if there is a matching of $G$ covering $Z$.  A
major strength of matroid theory comes from its two-fold feature:  on
one hand, it is pretty general to include an incredibly rich classes
of examples, on the other hand, it is rather special to allow
particularly deep theorems and algorithms.

As for matroidal generalizations of basic results of graph theory are
concerned, already the paper
\cite{Edmonds65a} of Edmonds contains the following extension of
Theorem \ref{Nash-fed} of Nash-Williams.

\THEOREM [Edmonds \cite{Edmonds65a}] \label{Edmonds-cover} The
ground-set of a matroid $M=(S,r)$ with rank function $r$ can be
partitioned into $k$ independent sets if and only if, for every subset $X\subseteq S$, \ $k\cdot
r(X)\geq \vert X\vert $. \  \eT

When $M$ is a graphic matroid, we are back at Nash-Williams' theorem, and when $M$
is a matroid representable over a field, we are back at Horn's Theorem
\ref{Horn}, up to some minor technical steps. This theorem immediately provides a
characterization of graphs whose edge set can be partitioned into $k$
forests, each consisting of at most $l$ edges.

However, a direct extension of Tutte's Theorem II to a matroidal
framework was described only in the fundamental paper
of Edmonds and Fulkerson \cite{Edmonds-Fulkerson}, who proved the
following result (formulated here in our terms).

\THEOREM [Edmonds and Fulkerson \cite{Edmonds-Fulkerson}, {\rm Theorem
2c}] \label{Edmonds+Fulkerson2c} Let $M_1,\dots ,M_k$ be matroids on a
common ground-set $S$.  There is a sub-partition $\{B_1,\ldots, B_k\}$
of $S$ for which $B_i$ is a basis of $M_i$ ($i=1,\dots ,k$) if and
only if, for every subset $X\subseteq S$,

$$\vert X\vert \geq \sum _{i=1}\sp k t_i(X),$$

\noindent   where $t_i$ denotes
the co-rank function of $M_i$ defined by $$ \hbox{ $t_i(X):  = \min \{
\vert X\cap B\vert :  B$ \ a basis of $M_i\}.$ }\ $$ \eT

Following this theorem, Edmonds and Fulkerson explicitly mention that in
the special case in which each $M_i$ is a graphic matroid, Theorem
\ref{Edmonds+Fulkerson2c} is nothing but Tutte's Theorem II (i.e. Theorem \ref{Tutte-pakol2}).

Edmonds and Fulkerson also proved a significant extension of Theorem
\ref{Edmonds-cover}, which may be viewed as the covering
counterpart of Theorem \ref{Edmonds+Fulkerson2c}.

\THEOREM [Edmonds and Fulkerson \cite{Edmonds-Fulkerson}, {\rm Theorem
1c}] \label{Edmonds+Fulkerson1c} Let $M_1,\dots ,M_k$ be matroids on a
common ground-set $S$.  There is a partition $\{I_1,\ldots, I_k\}$ of
$S$ for which $I_i$ is an independent set of $M_i$ ($i=1,\dots ,k$) if
and only if, for every subset $X\subseteq S$, 

$$\vert X\vert \leq \sum _{i=1}\sp k r_i(X),$$

\noindent where $r_i$ denotes
the rank function of $M_i$.  \eT

In the special case, when the matroids are graphic, this theorem
implies not only Theorem \ref{Nash-fed} but also its extension to the case in which, for given upper bounds $g_1,\dots ,g_k$, we want to find a
decomposition of $E$ into forests $F_1,\dots ,F_k$ for which $\vert
F_i\vert \leq g_i$ ($i=1,\dots ,k$).

It should be emphasized that there are constructive proofs of the two
theorems of Edmonds and Fulkerson which give rise to polynomial
algorithms to compute the $k$ independent sets or the $k$ bases in
question.

These general matroidal results of Edmonds and Fulkerson are extremely
helpful in properly understanding the background of the tree-covering
and tree-packing theorems, however the genuine recognition behind the real
power of the matroidal view is the following theorem of Edmonds from
1968.

\THEOREM [Matroid-sum theorem of Edmonds \cite{Edmonds68}, {\rm
Theorems 1 and 2}] \label{matroid-sum} Let $M_1 \allowbreak = (S,r_1),$ $ \dots ,
M_k=(S,r_k)$ be $k$ matroids on a common ground-set $S$, and let
${\cal F}_i$ denote the set of independent sets of $M_i$.  Let

\eq {\cal F}_\Sigma:= \{ F_1\cup \cdots \cup F_k \ :  \ F_i\in {\cal
F}_i \}.  \eeq

\noindent Then ${\cal F}_\Sigma$ forms the set of independent sets of
a matroid (called the sum or union of matroids $M_1,\dots ,M_k$).  The
rank function $r_\Sigma$ of $M_\Sigma$ is as follows.

\eq r_\Sigma(Z) = \min \{\vert Z-X\vert + \sum _{i=1}\sp k r_i(X) \ :
\ X\subseteq Z\}.  \eeq

\eT

Finally, it is useful to recall that Kruskal's algorithm \cite{Kruskal} has also been
extended to matroids to compute a maximum weight (or minimum cost)
basis of a matroid, see the paper \cite{Edmonds71} of Edmonds.

In this light, we can safely say that the theorems of Nash-Williams
and Tutte became the root and a major driving force of the huge area
of matroid optimization.

At this point, it is useful to remark that a fundamental feature of
matroids is that the rank-function of a matroid is submodular (that is, whenever $X,Y\subseteq S$$, r(X)+r(Y)\geq r(X\cap Y) + r(X\cup Y)$
).  We note that submodular functions (coming
from graphs) were already used by Tutte and Nash-Williams in their
proofs.

\subsection{Some consequences for graphs}

The matroidal background outlined above gave rise to solutions to the
first six problems listed in Section \ref{Problems}.  Here we mention
explicitly only one concrete extension of Theorem \ref{TNW} that can
be derived from the matroid results.

\THEOREM \label{FrankP4} For integers $k\geq 1$ and $h\geq 0$, a graph
$G=(V,E)$ can be made $k$-tree-connected by adding at most $h$ new
edges if and only if, for every partition $\cal P$ of $V$,

\eq e_G({\cal P}) \geq k\cdot (\vert {\cal P}\vert -1) - h.
\label{(Nash-pakol3)} \eeq 
\eT

By relying on the matroid-sum theorem of Edmonds, one can observe
that, for any spanning tree $(V,F)$ of the complete graph on $V$, the
$h$ new edges in the theorem whose addition to $G$ makes the graph
$k$-tree-connected can be chosen to be parallel to some edges of $F$.
In particular, the new edges may be chosen in such a way that they
form a star (or that they are parallel to some edges of a spanning
tree of $G$).

For a partition $\cal P$, the value $(k(\vert {\cal P}\vert -1)
-e_G({\cal P}))\sp +$ is called the {\bf $k$-deficit} of $\cal P$,
where $x\sp +:=\max\{0,x\}$.  We call the largest $k$-deficit over all
partitions of $V$ the {\bf $k$-partition-deficiency} of $G$ and denote
it by $\Pi_k(G)$.

Let us consider a partition $\cal P$ with maximum $k$-deficit.  It can
be shown that each node-set $V_i\in {\cal P}$ induces a
$k$-tree-connected graph, and the contraction of sets $V_i$ gives a
$k$-sparse graph.  Jackson and Jord\'an \cite{Jackson-Jordan10} showed
that there is a unique partition ${\cal P}\sp *$ among them with
$\vert {\cal P}\sp *\vert $ minimum (which they called the
brick-partition of $G$).  These observations give an alternative proof
of the following theorem.

\THEOREM The minimum number of new edges whose addition to a graph
$G=(V,E)$ results in a $k$-tree-connected graph is $\Pi_k(G)$.  The
maximum cardinality of the union of $k$ spanning trees of a connected
graph $G$ is $k(n-1) - \Pi_k(G)$.  \eT

To illustrate the progress in this area, we cite a recent theorem of Akrami, Raj and V\'egh \cite{ARV2025}
stating that if the ground-set $S$ of a matroid $M$ can be partitioned
into $k$ bases, then, for any subset $Z$ of $S$, there is a partition
$\{B_1,\dots ,B_k\}$ of $S$ into bases which is $Z$-equitable in the
sense that $\lfloor {\frac{\vert Z\vert}{ k}}\rfloor \leq \vert B_i\cap
Z\vert \leq \lceil {\frac{\vert Z\vert}{ k}}\rceil $ for $i=1,\dots ,k$.
For graphic matroids, this means that if a graph is decomposable into
$k$ spanning trees, then there is an equitable decomposition for any
subset of edges.

\section{Hypergraphs and directed graphs}

After exploring and understanding the matroidal background of the
tree-packing and tree-covering theorems, the question naturally
emerges:  how can the connectivity concepts and theorems concerning
undirected graphs be extended to hypergraphs, to directed graphs and
to dypergraphs ($=$ directed hypergraphs.)

\subsection{Hypergraphs}

Let $H=(V,{\cal E})$ be a hypergraph with node-set $V$ where $\cal E$
denotes the set of hyperedges.  We assume throughout that each
hyperedge has at least two elements.  By {\bf trimming} a hyperedge
$X$ to a graph-edge, we mean the operation of deleting $\vert X\vert
-2$ elements of $X$, or in other words, replacing $X$ with a
graph-edge connecting two elements of $X$.

We say that a hypergraph $H$ is a {\bf hyperforest} if its hyperedges
can be trimmed to a forest.  Lov\'asz \cite{Lovasz70a} proved that $H$
is a hyperforest if and only if the union of any $j>0$ hyperedges has
at least $j+1$ elements.  A {\bf spanning hypertree} is a hyperforest
having exactly $\vert V\vert -1$ hyperedges.

A natural extension of the notion of graph connectivity is as follows.
$H$ is {\bf connected} if $d_H(X)\geq 1$ for every non-empty proper
subset $X$ of $V$, where $d_H(X)$ is the number of hyperedges
intersecting both $X$ and $V-X$.  $H$ is {\bf partition-connected} if,
for every partition $\cal P$ of $V$, the number of hyperedges
intersecting at least two members of $\cal P$ is at least $\vert {\cal
P}\vert -1$.  While these two connectivity concepts are equivalent for
graphs, a hypergraph on a 3-element node-set $V$ with a single
hyperedge $V$ is connected but not partition-connected.

\THEOREM [Lorea \cite{Lorea75}] The hyperforests of a hypergraph $H =
(V,{\cal E})$ form the independent sets of a matroid on the ground-set
$\cal E$ of hyperedges.  \eT

A matroid arising in this way is called the {\bf hypergraphic} matroid
of $H$.

\THEOREM [Whiteley \cite{Whiteley96}] \label{hyper-rank} The rank of
the hypergraphic matroid of a hypergraph $ H=(V,{\cal E})$ is equal to

$$ \min \{ \vert V\vert - \vert {\cal P}\vert + e_H({\cal P}) :  {\cal
P} \hbox{ a partition of}\ V\} \label{(hyper-rank)} $$

\noindent where $e_H({\cal P})$ denotes the number of hyperedges
intersecting at least two members of the partition.  \eT

This implies that a hypergraph $H=(V,{\cal E})$ is partition-connected
if and only if the rank of its hypergraphic matroid is $\vert V\vert
-1$. The following theorem may be considered as a straight
generalization of the tree-packing and tree-covering theorems of Tutte
and Nash-Williams (Theorems \ref{TNW} and \ref{Nash-fed}) to
hypergraphs. A hypergraph $H$ is {\bf $k$-partition-connected} if,
for every partition $\cal P$ of $V$, the number of hyperedges
intersecting at least two members of $\cal P$ is at least $k(\vert {\cal
P}\vert -1)$.

\THEOREM [Frank, Kir\'aly, Kriesell \cite{FrankJ49}] Let $H=(V,{\cal
E})$ be a hypergraph.

{\rm {\bf (A)} } \ $\cal E$ can be decomposed into $k$ spanning
partition-connected subhypergraphs (or, equivalently, $H$ includes $k$
disjoint spanning hypertrees) if and only if $H$ is
$k$-partition-connected.

{\rm {\bf (B)} } \ $\cal E$ can be decomposed into $k$ hyperforests if
and only if we have

$$ i_{\cal E}(X) \leq k\cdot (\vert X\vert -1) $$ for every non-empty
subset $X$ of $V$, where $i_{\cal E}(X)$ denotes the number of
hyperedges included in $X$.  \eT

This theorem provides a solution to Problem \ref{hypergraph}.  As for
algorithmic aspects, see the paper \cite{Baiou+Barahona} of Ba{\"\i}ou
and Barahona.  Note that the problem of decomposing a hypergraph into
$k$ connected hypergraphs is already {\bf NP}-hard for $k=2$
\cite{FrankJ49}.

\subsection{Directed and mixed graphs}

A directed tree is an {\bf arborescence} or {\bf $r_0$-arborescence}
if it has a node $r_0$ of in-degree 0 (called the root-node) and the
in-degree of every other node is one.  The union of node-disjoint
arborescences is a {\bf branching}.  A digraph $D=(V,A)$ is called
{\bf root-connected} (from a root-node $r_0\in V$) if the in-degree
$\varrho _D(X)$ of every non-empty set $X\subseteq V-r_0$ is positive.
This is equivalent to requiring that $D$ has a spanning
$r_0$-arborescence, or that every node of $D$ is reachable from $r_0$
along a directed path.  $D$ is {\bf rooted $k$-arc-connected} if
$\varrho _D(X)\geq k$ for every non-empty set $X\subseteq V-r_0$.
This is equivalent (by Menger) to requiring that there are $k$
arc-disjoint $r_0v$-dipaths for each $v\in V-r_0$.

The following fundamental theorem of Edmonds may be viewed as the
directed counterpart of the tree-packing theorem of Tutte and
Nash-Williams.

\THEOREM [Arborescence-packing theorem of Edmonds \cite{Edmonds73}]
\label{disjoint-arb} {\rm (Weak form)} \ A digraph $D=(V,A)$ with a
root-node $r_0$ contains $k$ arc-disjoint spanning $r_0$-arborescences
if and only if $D$ is rooted $k$-arc-connected.  {\rm (Strong form)} \
Let $(V,A_1),\dots ,(V,A_k)$ be $k$ arc-disjoint $r_0$-arborescences
of $D$, and let $A_0:= A- (A_1\cup \cdots \cup A_k)$.  They can be
extended to $k$ arc-disjoint spanning arborescences if and only if
$\varrho _{A_0}(X)$ plus the number of arborescences $A_i$ entering
$X$ is at least $k$ for every non-empty subset $X\subseteq V-r_0$.
\eT

Lov\'asz \cite{Lovasz76a} found a stunningly short and simple proof of
this theorem, but a major difference between the directed and the
undirected cases is that, unlike the tree-packing theorem of Tutte and
Nash-Williams, no matroidal result is known that implies Edmonds'
theorem.

We show how Edmonds' theorem helps solve Problem
\ref{min-cost-digraph1}.  Let $c$ be a non-negative cost-function on
the arc-set of an $r_0$-rooted $k$-arc-connected digraph $D=(V,A)$
with no arc entering $r_0$.  Let $G=(V,E)$ denote the undirected graph
obtained from $D$ by de-orienting each arc.  Let $M_1$ be the matroid
on $A$ in which a set of arcs is a basis of the corresponding set of
edges in $G$ is the union of $k$ disjoint trees (that is, $M_1$ may be
viewed as $k$ times the graphic matroid of $G$).  Let $M_2$ be a
partition matroid on $A$ in which a subset of arcs is a basis if the
in-degree of every non-root node is $k$.  By relying on Theorem
\ref{disjoint-arb} of Edmonds, it is not difficult to see for a subset
$J\subseteq A$ of arcs that the subgraph $(V,J)$ of $D$ is a minimally
(with respect to arc deletion) rooted $k$-arc-connected digraph if and
only of it is a common basis of $M_1$ and $M_2$.  Therefore an
algorithm for computing a minimum cost common basis of two matroids
can be applied \cite{Edmonds79}.

An essential difference between the directed and the undirected cases
is that, while there is an efficient algorithm (via matroids) for
solving Problem \ref{multi-cost}, its directed counterpart in Problem
\ref{multi-cost-digraph} is NP-hard already for $k=2$.

\medskip

As far as the directed counterpart of Nash-Williams tree-covering
theorem is concerned, there are two results.

\THEOREM [Vidyasankar \cite{Vidyasankar}] Let $r_0$ be a root-node of
a digraph $D=(V,A)$ such that no arc enters $r_0$.  The arc-set of $D$
can be covered by $k$ spanning $r_0$-arborescences if and only if {\rm {\bf (a)} } \ $\varrho _D(v) \leq k$ for every $v\in V - r_0$ and
\ {\rm {\bf (b)} } \ $k - \varrho _D(X) \leq \sum [k -\varrho _D(v) :
v \in \Gamma \sp -(X)]$ for every non-empty set $X\subseteq V-
r_0$, \
 where $\Gamma \sp -(X)$ denotes the set of heads of the arcs
entering $X$.  \eT

\THEOREM [Frank \cite{FrankP4}] The arc-set of a digraph $D = (V, A)$
can be covered by $k$ branchings if and only if  \ {\rm {\bf (a)} } \
the in-degree of each node is at most $k$ and \ {\rm {\bf (b)} } \ $
i_D(X) \leq k\cdot (\vert X\vert - 1)$ for every set $\emptyset
\subset X\subseteq V$, where $i_D(X)$ denotes the number of arcs
induced by $X$.  \eT

It is possible to merge the tree-packing and the arborescence-packing
theorems.  A {\bf mixed graph} $M=(V, A+E)$ is a graph with possible
directed and undirected edges.  A mixed $r_0$-arborescence is a mixed
tree in which it is possible to orient the undirected edges to get 
an arborescence \cite{FrankP4}.

\THEOREM [\cite{FrankP4}] In a mixed graph $M=(V, A+E)$ with root-node
$r_0$, there are $k$ edge-disjoint spanning mixed $r_0$-arborescences
if and only if, for every partition ${\cal
P}=\{V_0,V_1,\dots ,V_q\}$ of $V$, $e_G({\cal P}) \geq \sum [k-\varrho _A(V_i):  \
i=1,\dots ,q ],$ \  where $r_0\in V_0$ and $e_G({\cal
P})$ denotes the number of edges from $E$ connecting distinct parts of
$\cal P$.  \eT

Note that these results provide an answer to Problem \ref{digraph1}.

\subsection{Dypergraphs}

Analogously to digraphs, one may consider directed hypergraphs or
dypergraphs.  A {\bf dyperedge} (directed hyperedge) $(Z,z)$ is a pair consisting 
of a set $Z$ with $\vert Z\vert \geq 2$ and an element $z$ of $Z$,
where $z$ is called the {\bf head} of $Z$ while the remaining elements
of $Z$ are its {\bf tails}.  A dyperedge $(Z,z)$ is said to enter a
set $X\subset V$ if $z\in X$ and $Z-X$ is non-empty.  A dypergraph $D
=(V, {\cal A})$ consists of a node-set $V$ and a family ${\cal A}$ of
dyperedges.  The in-degree $\varrho _D(X)$ of a subset $X\subseteq V$
is the number of dyperedges entering $X$.

Given a root-node $r_0$, we say that a dypergraph is {\bf out-rooted
$k$-arc-connected} if the in-degree of every non-empty set $X\subseteq
V-r_0$ is at least $k$.  When $k=1$, $D$ is out-rooted arc-connected.
The dypergraphic extension of Edmonds' arborescence-packing theorem is
as follows.

\THEOREM [\cite{FrankJ49}] Suppose that every dyperedge of a
dypergraph $D = (V, {\cal A})$ has at least two elements.  Let $r_0\in
V$ be a given root-node.  Then $D$ can be decomposed into $k$
out-rooted arc-connected dypergraphs if and only if $D$ is out-rooted
$k$-arc-connected.  \eT

\section{Orienting graphs and hypergraphs}

\subsection{Graphs}

Orienting an undirected edge $e=uv$ means that we replace $e$ with one
of the two directed edges ($=$ arcs) $uv$ and $vu$.  Orienting a
hyperedge $Z$ means that we replace $Z$ by a dyperedge $(Z,z)$ for
some $z\in Z$.

Graph (hypergraph) orientation problems serve as a bridge between
directed graphs and undirected graphs (hypergraphs).  For example, the
weak form of Nash-Williams' orientation theorem \cite{NWir} states
that a graph has a $k$-arc-connected orientation if and only if $G$ is
$2k$-edge-connected.  A general graph orientation result
\cite{FrankJ4} implies the following.

\THEOREM \label{root-con-ori} A graph $G$ has a rooted
$k$-arc-connected orientation if and only if $G$ is
$k$-partition-connected.  \eT

On the one hand, if $G$ is $k$-partition-connected, then, by Theorem
\ref{TNW}, there are $k$ disjoint spanning trees, and we can orient
them separately to obtain $k$ disjoint spanning arborescences.  On the
other hand, this orientation theorem and the disjoint arborescence
theorem of Edmonds immediately implies Theorem \ref{TNW}.

For non-negative integers $k$ and $l$, a digraph or a dypergraph $D$
on node-set $V$ with a specified root-node $r_0\in V$ is {\bf rooted
$(k,l)$-arc-connected} if $\varrho _D(X)\geq k$ and $\varrho
_D(V-X)\geq l$ whenever $\emptyset \subset X\subseteq (V-r_0)$.

\THEOREM [\cite{FrankJ4}] \label{(k,l)-ori} For non-negative integers
$k$ and $l$, an undirected graph $G$ with a root-node $r_0$ has a
rooted $(k,l)$-arc-connected orientation if and only if $G$ is
$(k,l)$-partition-connected.  \eT

\medskip

Incidentally, we note that the concept of $k$-edge-connec\-ti\-vi\-ty is
simpler than $k$-tree-connec\-ti\-vi\-ty.  However, if we want a concise
certificate for a graph to be $k$-edge-connected, then the simplest
one (so far) is a set of $k$ edge-disjoint $r_0v$ paths for every node
$v\in V-r_0$.  But there is a more concise certificate implied by Theorem \ref{root-con-ori}.

\THEOREM A graph $G=(V,E)$ is $k$-edge-connected if and only if, for
an arbitrarily specified node $r_0\in V$, the digraph $D=(V,A)$
arising from $G$ by replacing each edge by two oppositely directed
parallel edges contains $k$ disjoint spanning arborescences of root
$r_0$.  \eT

\subsection{Hypegraphs}

As for orientation problems of hypergraphs are concerned, one has the
following.

\THEOREM [\cite{FrankJ49}] \label{k-hyperarc} {\rm {\bf (A)} } \ A
hypergraph $H=(V,{\cal E})$ has a $k$-arc-connected orientation if and
only if, for
every partition $\cal P$ of $V$, $e_H({\cal P}) \geq k \cdot \vert {\cal P}\vert, $  where $e_H({\cal P})$ denotes the
number of hyperedges intersecting at least two members of $\cal P$.
{\rm {\bf (B)} } $H$ has an out-rooted $k$-arc-connected orientation
(with respect to a specified root-node) if and only if $H$ is
$k$-partition-connected.  \eT

A trivial observation is that a digraph is rooted
$(k,l)$-arc-connected if and only if the digraph obtained by reversing
all arcs is $(l,k)$-arc-connected.  Because of the asymmetric role of
tails and heads in a dypergraph, this observation does not extend to
dypergraphs.  For example a dypergraph on node-set $V:=\{r_0,s,t\}$
having the single hyperedge $V$ has a $(0,1)$-partition-connected
orientation but it has no $(1,0)$-partition-connected orientation.

In a paper of Guo, Li, Shangguan, Tamo,
and Wootters~\cite{GLSTW2024}, a hypergraph $H$ is called
{\bf $h$-weakly-partition-connected} if $\sum [d_{\cal P}(Z) -1:  \
Z\in {\cal E}] \geq h \cdot ( \vert {\cal P}\vert -1)$ for every
partition $\cal P$ of $V$, where $d_{\cal P}(Z)$ denotes the number of
members of $\cal P$ intersecting $Z$.  By relying on this
concept, they formulated a general conjecture concerning Reed-Solomon
codes \cite{R-S}.  The question whether there is a good characterization of this
property is answered in a more general form by the following result.

\THEOREM [\cite{FrankJ50}] \label{hyper-ori} Let $H = (V, {\cal E})$
be a hypergraph with a root-node $r_0$, and let $k$ and $l$ be
non-negative integers.

{\rm {\bf (A)} } In the case $l\leq k$, $H$ has a $(k,l)$-arc-connected
orientation if and only if, for every partition $\cal P$ of $V$,

\eq e_H ({\cal P}) \geq k\cdot (\vert {\cal P}\vert - 1) + l.
\label{(hyper-oriA)} \eeq

\noindent  In the special
cases when $l=k$ or $l=0$, we are back at Parts (A) and (B),
respectively, of Theorem \emref{k-hyperarc}.

{\rm {\bf (B)} } In the case $l>k$, $H$ has a $(k,l)$-arc-connected
orientation if and only if, for every partition $\cal P$ of $V$, \emeref{(hyper-oriA)} holds and

\eq \sum _{Z\in {\cal E}} [ d_{\cal P}(Z) -1 ] \geq l \cdot (\vert
{\cal P}\vert - 1) + k, \label{(hyper-oriB)} \eeq

\noindent  where $d_{\cal
P}(Z)$ denotes the number of members of $\cal P$ intersecting $Z$.  In the special case in which $k=0$, we obtain that $H$ has an orientation in which the out-degree of every non-empty subset
of $V - r_0$ is at least $l$ if and only if $H$ is
$l$-weakly-partition-connected.  \eT

We remark that the special case $k=0$ of Part (B) (for characterizing
$l$-weakly-partition-connected hypergraphs) was applied in a recent
paper of Alrabiah, Guo, Guruswami, Li, and Zhang~\cite{AGGLZ2025} in the theory of
Reed-Solomon codes, where the in-degrees of such orientations were used to create generic zero patterns.

The proof of Theorem \ref{hyper-ori} gives rise to a polynomial
algorithm that finds either a $(k,l)$-arc-connected orientation or a
partition $\cal P$ violating the conditions.

\section{Constructive characterizations}

By a constructive characterization of a graph with a certain property,
we mean a simple construction to build up the graph with the given
property step by step.  For example, a graph is connected if and only
if it can be built up from a node by adding consecutively new edges
for which at least one end-node is an already existing node.
Constructive characterizations may be useful for deriving theorems
concerning the graph property in question.
The constructions presented here apply new node/edge/arc additions or \textbf{pinching} some existing edges (arcs) with a new node, which means that we subdivide first some edges (arcs) by new nodes and then merge these new nodes into a single one.
In his proof of Theorem \ref{TNW}, Nash-Williams also used this pinching technique to (implicitly) give a constructive characterization of $k$-forest-sparse graphs with $k|V|-k$ edges.

In this section we briefly list some fundamental constructive characterizations concerning
various graph connectivity concepts. For example, Theorem \ref{k-part-con-constr} below provides such a
characterization of $k$-partition-connected graphs, and this result immediately implies Theorem
\ref{TNW}.

\THEOREM [Lov\'asz \cite{Lovasz}] An undirected graph is
$2k$-edge-connected if and only if it can be built up from an initial
node by consecutively applying the following two operations.

\noindent (A) \ Add a new edge (possibly a loop) connecting two
existing nodes.

\noindent (B) \ Pinch together $k$ existing edges with a new node.
\eT

Note that this constructive characterization immediately implies a
theorem of Nash-Williams stating that a $2k$-edge-connected graph has
a $k$-arc-connected orientation.

\THEOREM [Mader \cite{Mader82}] \label{Mader-splitting} A digraph is
$k$-arc-connected if and only if it can be built up from an initial
node by consecutively applying the following two operations.

\noindent (A) \ Add a new arc (possibly a loop) connecting two
existing nodes.

\noindent (B) \ Pinch together $k$ existing arcs with a new node.  \eT

This result rather easily implies the following.

\THEOREM \label{root-con-constr} A digraph $D=(V,A)$ with a root-node
$r_0$ is rooted $k$-arc-connected if and only if it can be built up
from $r_0$ by consecutively applying the following two operations.

\noindent (A) \ Add a new arc (possibly a loop) connecting two
existing nodes.

\noindent (B) \ Add a new node $z$ along with $k$ arcs entering $z$
for which their tails are existing nodes.

\noindent (C) \ Pinch together $j$ \ $(0<j\leq k)$ existing arcs with
a new node $z$ and add $k-j$ new arcs entering $z$ whose tails are
existing nodes.  \eT

By combining Theorems \ref{root-con-constr} and \ref{root-con-ori},
one obtains the following.

\THEOREM \label{k-part-con-constr} An undirected graph $G$ is $k$-partition-connected if and
only if it can be built up from a node by consecutively applying the
following operations.

\noindent (A)\ Add a new edge (possibly a loop) connecting two
existing nodes.

\noindent (B) \ Add a new node $z$ along with $k$ edges connecting $z$
with existing nodes.

\noindent (C)\ Pinch together $j$ existing edges ($0< j\leq k$) with a
new node $z$ and add $k-j$ (possibly parallel) new edges connecting
$z$ with existing nodes.  \eT

This result rather easily implies the tree-packing theorem of Tutte
and Nash-Williams (Theorem \ref{TNW}).  A significant extension of
Theorem \ref{root-con-constr} is as follows.

\THEOREM [Kov\'acs and V\'egh \cite{Kovacs-Vegh10}]
\label{Kovacs+Vegh} Let $0\leq l\leq k-1$ be integers.  A digraph $D$
is $(k,l)$-arc-connected with respect to a root-node $r_0$ if and only
if $D$ can be built up from $r_0$ by consecutively applying the
following operations.

\noindent (A)\ Add a new arc connecting two existing nodes.

\noindent (B)\ Pinch together $j$ existing arcs ($l\leq j\leq k-1$)
with a new node $z$ and add $k-j$ (possibly parallel) new arcs
entering $z$ for which their tails are existing nodes.\eT

By combining Theorems \ref{(k,l)-ori} and \ref{Kovacs+Vegh}, one gets the following result which shows how to construct (all) graphs that include $k$ disjoint spanning trees after removing any set of $l$
edges.

\THEOREM \label{(k,l)-part-con-constr} Let $0\leq l< k$ be integers.
An undirected graph $G$ is $(k,l)$-partition-connected if and only if
it can be built up from a node by consecutively applying the following operations.

\noindent (A)\ Add a new edge (possibly a loop) connecting two
existing nodes.

\noindent (B)\ Pinch together $j$ existing edges ($l\leq j\leq k$)
with a new node $z$ and add $k-j$ (possibly parallel) new edges
connecting $z$ with existing nodes.  \eT

The proof of this theorem is deep and difficult, however, a
significantly simpler proof is available for its special case $l=1$ in
the paper of Frank and Szeg{\H o} \cite{FrankJ48}.  This special case
has an interesting application in rigidity theory, see Section \ref{rigidity}.

\medskip


\section{Tree-connectivity in rigidity theory}\label{rigidity}

Tree-connectivity finds applications in seemingly distant areas such
as rigidity theory.  Rigidity theory in this context examines the stability of structures
modeled as graphs embedded in the Euclidean space with straight-line
edges.  The primary goal is to determine whether these structures
(called frameworks) can be deformed without changing the edge lengths.
If such a deformation does not exist, the framework is called rigid.
In a bar-and-joint framework, nodes of a graph (the joints) are
embedded in a $d$-dimensional space, where the edges (the bars) have
fixed lengths.  Intuitively, such a framework is locally rigid - rigid, for short- if a continuous deformation of the framework preserves pairwise distances of the nodes.  An embedding of a graph is generic if all coordinates of
the node positions are algebraically independent over the rational
numbers.  Rigidity is a generic property \cite{Whiteley96}, and in
some cases, generic rigid graphs can be characterized by sparsity.
Here we call a graph $G=(V,E)$ $(k,l)$-forest-tight if it is
$(k,l)$-forest-sparse and $\vert E\vert = k\cdot (\vert V\vert -1)-l$.
Laman \cite{Laman} proved that $(2,1)$-forest-tight graphs are the
generic minimally rigid graphs for bar-and-joint frameworks in the
plane, and a graph is rigid in the plane if and only if it contains a
$(2,1)$-forest-tight subgraph.

A body-and-bar framework is similar to a bar-and-joint framework,
except that the nodes of a graph represent full-dimensional bodies that are
connected by fixed-length bars such that the attachment points of the
bars on the bodies are pairwise distinct.  Tay showed \cite{Tay1984}
that a multi-graph (a graph with possibly parallel edges but no loops)
can be realized by a rigid body-bar framework in dimension $d$ if and
only if it contains $d(d+1)/2$ edge-disjoint spanning trees.
Informally, a framework is globally rigid if for every other embedding
with the same bar lengths also the distances of every pair of nodes
are the same.  A graph $G$ is highly $k$-tree-connected if $e_G({\cal
P}) \geq k\cdot (\vert {\cal P}\vert - 1) + 1$.  There is also a
strong relation between generic global rigidity of body-bar frameworks
and highly tree-connected graphs.  A framework is called redundantly
rigid if it is rigid and remains rigid after the removal of any edge.
For a multi-graph $H$, a so-called body-bar graph $G_H$ can be
associated with $H$.  (Its definition can be found for example in
\cite{CJW}.)

Connelly et al.  \cite{CJW} showed that the following are equivalent:
(a) $G_H$ is generically globally rigid in ${\bf R} \sp d$, (b) $G_H$
is generically redundantly rigid in ${\bf R} \sp d$, (c) $H$ is highly
$\binom{d+1}{2}$-tree-connected.  Their proof relies on the
constructive characterization of highly $k$-tree-connected graphs,
that is, $(k,1)$-partition-connected graphs (see Theorem
\ref{(k,l)-part-con-constr} for the special case $l=1$).  We note that
generic globally rigid body-hinge frameworks can also be characterized
by highly tree-connected graphs \cite{JKT2016}.


We saw above that connectivity and constructive characterizations are
closely related, and the latter turned out to be a useful tool in
rigidity.  Interestingly, this connection is bidirectional:  recent
results in rigidity were the key to prove some long-standing
conjectures about connectivity.  Kriesell conjectured that a graph
with high node-connectivity can always be partitioned into a spanning
tree and a $k$-connected subgraph \cite{MNW2007}.  His conjecture has
recently been proven by Garamv\"olgyi et al.  \cite{GJKW2025}.  They
actually prove a stronger statement about subgraphs with higher
dimensional rigidity.

\THEOREM [Garamv\"olgyi, Jord\'an, Kir\'aly, Vill\'anyi  \cite{GJKW2025}] Every $(t \cdot 10d(d
+ 1))$-connected graph contains $t$ edge-disjoint $d$-rigid (and hence
$d$-connected) spanning subgraphs.  \eT

They use a probabilistic approach in their proof.  We note that with
this tool they also answer an old conjecture of Thomassen
\cite{Thomassen1989} about highly connected orientations.

\THEOREM [Garamv\"olgyi, Jord\'an, Kir\'aly, Vill\'anyi  \cite{GJKW2025}] Every $(320 \cdot
k\sp 2)$-connec\-ted graph has a $k$-connec\-ted orientation.  \eT

For a more detailed description of these recent results see
\cite{CJJT2025}.

\section{Switching games}

Another, somewhat surprising application of disjoint spanning trees is the Switching game invented by Shannon~ (see \cite{Gardner}). 
He introduced a two-person game on connected
undirected graphs with two specified nodes $s$ and $t$.  The players,
Short and Cut, alternately tag an untagged edge, as long as there are
untagged edges.  Short wins if she tagged all edges of an $st$-path,
and Cut wins if she tagged all edges of a cut separating $s$ and $t$.

A special case of the switching game is the Bridge-It game of Gale
played on a square grid \cite{Gale2009}.  A matroidal extension is due
to Lehman \cite{Lehman1964} while Edmonds \cite{Edmonds65b} described
a general framework to explore the link between Lehman's results and
the theorem \ref{TNW} of Nash-Williams and Tutte.

For technical simplicity, here we discuss a version of Shannon's
switching game, where no specified nodes $s$ and $t$ are designated. This version shows the connection to tree-connectivity more directly, but it can also be deduced from the original game \cite{Lehman1964}.
Short wins if she tagged all the edges of a spanning tree of $G$ and
Cut wins if she tagged all the edges of a cut of $G$.  Clearly, at
most one of the two players can win, and an immediate observation
shows that one of them will definitely win.

Actually, there are two variants of the game, depending on who is the
first player.  Since these two variants are close to each other, we
assume that Cut tags first. 

\THEOREM Suppose that $G=(V,E)$ is a connected graph with $\vert
V\vert \geq 3$.

\noindent {\rm {\bf (A)} } There is a winning strategy for Short if
and only if $G$ is 2-tree-connected.

\noindent {\rm {\bf (B)} } There is a winning strategy for Cut if and
only if $V$ has a deficient partition ${\cal P}=\{V_1,\dots ,V_q\}$
for $k=2$, that is, one for which $e_G({\cal P}) < 2q-2.$ \eT

\Proof (outline) By Theorem \ref{TNW}, $G$ is 2-tree-connected if and
only if there is no deficient partition.  Therefore, exactly one of
the two configurations can occur.

If $F_1$ and $F_2$ are two disjoint spanning trees, and Cut tags an
edge $e$, say, from $F_1$, then Short should tag an edge $f\in F_2$
that connects the two components of $F_1-e$.  (If $e$ is not in $F_1\cup
F_2$, then any edge $f$ will do).  Now the graph $G'$ arising from
$G-e$ by contracting $f$ is also 2-tree-connected, and by using
iteratively this approach, Short can win.

If there is a deficient partition ${\cal P}$, a winning strategy for
Cut is that she always tags an untagged cross-edge of $\cal P$.  After
her first move, there are at most $2q-4$ untagged cross-edges.  During
the entire game Short can tag at most half of these edges, but every
spanning tree contains at least $q-1$ cross-edges, therefore Short
will loose, and Cut wins.  \FBOX

\medskip It is not difficult to formulate an analogous theorem for the
case when Short tags first.  Roughly, in this case, there is a winning
strategy for Cut if and only if there is a partition with $2$-deficit
at least 2 (which is equivalent to requiring that every two spanning
trees have at least two edges in common), and there is a winning
strategy for Short if and only if there are two spanning trees with at
most one edge in common (which is equivalent to requiring that the
2-deficiency of $G$ is at most 1).

For the original Shannon switching game (for cutting or connecting $s$
and $t$) when Cut is again the first player, one has the following.

\THEOREM \label{eredeti-Shannon} \noindent {\rm {\bf (A)} } There is a
winning strategy for Short if and only if there exists a subset $U$ of
nodes containing $s$ and $t$ for which $G\vert U$ is 2-tree-connected,
where $G\vert U$ denotes the restriction of $G$ to $U$.

\noindent {\rm {\bf (B)} } There is a winning strategy for Cut if and
only if $V$ has a partition ${\cal P}=\{V_1,\dots ,V_q\}$ separating
$s$ and $t$ such that the graph arising from $G$ by contracting each
$V_i$ to a single node can be decomposed into two forests $F_1$ and
$F_2$ where $s$ and $t$ belong to distinct components of $F_1$.  \eT

This theorem is a consequence of Edmonds' general
result \cite{Edmonds65b}  concerning games on matroids.  From an algorithmic aspect, it
is useful mentioning that if the unique maximum $2$-deficit partition
${\cal P}\sp *$ ensured by Jackson and Jord\'an
\cite{Jackson-Jordan10} (that is the brick partition of $G$) happens
to separate $s$ and $t$, then ${\cal P}\sp *$ will meet the
requirements in Part (B).

Theorem \ref{eredeti-Shannon} gives a characterization of the winning
cases of player Cut when she starts the game.  However, the result
itself does not provide an algorithm for deciding whether property (A)
or (B) holds.  It can be decided by finding a pair of maximally
disjoint spanning trees \cite{Baron+Imrich1968}.  The approach of
Bruno and Weinberg \cite{Bruno+Weinberg1970} uses the structural
result of Kishi and Kajitani \cite{Kishi+Kajetani1969} on maximally
disjoint trees, which is motivated by an application in electrical
networks, where the above distances of the spanning trees correspond
to certain degrees of freedom of a given electric network
\cite{Kron1939, Recskibook}.



\begin{thebibliography}{999}




\bibitem{ARV2025}
H.  Akrami, R.  Raj, and
L.A. V\'egh, 
{\it Matroids are equitable}
, Proceedings of the 2026 Annual ACM-SIAM Symposium on Discrete Algorithms (SODA), Society for Industrial and Applied Mathematics, (2026) 5843-5860.

\bibitem{AGGLZ2025}
O. Alrabiah, Z. Guo, V. Guruswami, R. Li, and Z.
Zhang,
{\it Random Reed-Solomon codes achieve list-decoding capacity with
linear-sized alphabets}, Advances in Combinatorics, 8 (2025).




\bibitem{Baiou+Barahona} M. Ba{\"\i}ou and F. Barahona,
{\it An algorithm for packing hypertrees}, Discrete Mathematics, 348
(2025) 114397.





\bibitem{Baron+Imrich1968}
G. Baron and W. Imrich, {\it On the maximal distance of spanning
trees}, Journal of Combinatorial Theory, 5(4) (1968) 378-85.


\bibitem{Bruno+Weinberg1970}
J. Bruno and L. Weinberg, {\it A constructive graph-theoretic solution
of the Shannon switching game}, IEEE Transactions on Circuit
Theory, 17 (1) (1970) 74-81.



\bibitem{Hall-thm}
P.J. Cameron, {\it Hall's marriage theorem}, The Journal of the London Mathematical
Society, (2026) 113: e70378.

 \bibitem{CJW}
R. Connelly, T. Jord\'an, and W. Whiteley, {\it Generic global
rigidity of body-bar frameworks}, J. Comb.  Theory, Ser.  B, 103(6)
(2013) 689-705.



\bibitem{CJJT2025}
J. Cruickshank, B. Jackson, T. Jord\'an,  and S. Tanigawa, {\it
Rigidity of graphs and frameworks:  a matroid theoretic approach},
arXiv preprint arXiv:2508.11636. (2025).


\bibitem{Dijkstra}
E. W. Dijkstra, {\it A note on two problems in connexion with graphs}, Numerische Mathematik. 1 (1959) 269–271. 

\bibitem{Dilworth}
R. P. Dilworth, {\it A decomposition theorem for partially ordered sets}, Annals of Mathematics, 51 (1)   (1950) 161–166. 




\bibitem{Edmonds65a}
J. Edmonds, {\it Minimum partition of a matroid
into independent sets,} J. Res.  Nat.  Bur.  Standards, B69
(1965) 67-72.

\bibitem{Edmonds65b} J. Edmonds, {\it Lehman's switching game and a
theorem of Tutte and Nash-Williams,} Journal of Research of the National Bureau of Standards,
B69 (1965) 73-77.

\bibitem{Edmonds68}
J. Edmonds, {\it Matroid Partition}, Mathematics of the Decision
Sciences, Part I. ) G.B.  Dantzig and A.F.  Veinott, eds.), American
Mathematical Society, (1968) 335-345.




\bibitem{Edmonds71}
J. Edmonds, {\it Matroids and the greedy
algorithm}, Math. Programming, 1 (1971) 127-136.


\bibitem{Edmonds73}
 J. Edmonds, {\it Edge-disjoint branchings,} in:
Combinatorial Algorithms (B.  Rustin, ed.), Acad.  Press, New York,
(1973) 91-96.


\bibitem{Edmonds79}
J. Edmonds, {\it Matroid intersection}, Annals of
Discrete Math. 4, (1979) 39-49.

\bibitem{Edmonds-Fulkerson} J. Edmonds and D.R.  Fulkerson, {\it
Transversals and matroid partition}, Journal of Research of the
National Bureau of Standards, B69 (1965)  147-153.

\bibitem{Frank-book}
A. Frank, Connections in Combinatorial
Optimization, \ Oxford University Press, 2011,  Oxford Lecture Series in Mathematics and its
Applications, 38.


\bibitem{FrankJ4}
Frank, {\it On the orientation of graphs}, Journal of 
Combinatorial Theory, Ser. B, Vol. 28(3) (1980) 251-261.


\bibitem{FrankP4}
A. Frank, {\it On disjoint trees and arborescences,} Algebraic Methods in Graph Theory, Colloquia Mathematica Soc.  J.
Bolyai, 25 (1981) 159-169.  North-Holland.  (Conference held at Szeged, Hungary, 1978)



\bibitem{FrankJ49}
A. Frank, T. Kir\'aly, and M. Kriesell, {\it On
decomposing a hypergraph into $k$ connected sub-hypergraphs,} in:
Submodularity, Discrete Applied
Mathematics, 131 (2) (2003) 373-383.


\bibitem{FrankJ50}
 A. Frank, T. Kir\'aly, and Z. Kir\'aly, {\it On the
orientation of graphs and hypergraphs,} in:  Submodularity, Discrete Applied Mathematics, 131 (2)
(2003) 385-400.


\bibitem{FrankJ48}
A. Frank and L. Szeg{\H o}, {\it Constructive
characterizations for packing and covering with trees,}
Submodularity, (guest editor S. Fujishige) Discrete Applied
Mathematics, 131(2) (2003)  347-371.






\bibitem{Gale2009}
D. Gale, {\it Topological games at Princeton, a mathematical memoir},
Games and Economic Behavior,  66(2) (2009) 647-56.





\bibitem{Gardner}
M. Gardner, The Second Scientific
American Book of Mathematical Puzzles and  Diversions, The University
of Chicago Press, (1961).


\bibitem{GJKW2025}
D. Garamv\"olgyi, T. Jord\'an, Cs.  Kir\'aly, and S. Vill\'anyi, {\it
Highly connected orientations from edge-disjoint rigid subgraphs}, In:
Forum of Mathematics, Pi, 13 (2025) 1-15, Cambridge
University Press.

\bibitem{GareyJohnson}
M.R. Garey, D.S. Johnson, Computers and Intractability: A Guide to the Theory of NP-Completeness. W. H. Freeman and Company, (1979).

\bibitem{GLSTW2024}
Z. Guo, R. Li, C. Shangguan, I. Tamo,
and M. Wootters, {\it Improved list-decodability and
list-recoverability of Reed-Solomon codes via tree packings}, SIAM
Journal of Computing, 53(2) (2024).




\bibitem{Horn55} A. Horn, {\it A
characterization of unions
of linearly independent sets}, The Journal of the London Mathematical
Society, 30(4) (1955)
494--496.





\bibitem{Jackson-Jordan10}
 B. Jackson and  T. Jord\'an, {\it Brick
partitions of graphs}, Discrete Mathematics, 310(2)
(2010) 270-275.

\bibitem{JKT2016}
T. Jord\'an, Cs.  Kir\'aly, and S. Tanigawa, {\it Generic global
rigidity of body-hinge frameworks}, Journal of Combinatorial Theory,
Series B,  117 (2016) 59- 76.



\bibitem{Karger2000}
D.R.  Karger, {\it Minimum cuts in near-linear time}, Journal of the
ACM (JACM), 47(1) (2000) 46--76.


\bibitem{Kishi+Kajetani1969}
G. Kishi and Y. Kajitani, {\it Maximally distant trees and principal
partition of a linear graph}, IEEE Transactions on Circuit Theory.
16(3) (1969) 323--330.


\bibitem{Kovacs-Vegh10}
R.E.  Kov\'acs and L.A.  V\'egh, {\it
Constructive characterization theorems in combinatorial
optimization,}
Combinatorial Optimization and Discrete Algorithms (ed.  S.
Iwata), RIMS Kokyuroku Bessatsu B23, (2010) 147--169.

\bibitem{Kron1939}
G. Kron,  Tensor Analysis of Networks,  New York: J. Wiley \& Sons;
(1939).

\bibitem{Kruskal}
J. Kruskal, {\it On the shortest spanning subtree and the traveling salesman problem},   Proceedings of the American Mathematical Society, 7, (1956), 48--50.



\bibitem{Laman}
G. Laman, {\em On graphs and rigidity of plane skeletal structures},
Journal of Engineering Mathematics, 4 (1970) 331--340.

\bibitem{Lee+Streinu2008}
A. Lee and I. Streinu, {\it  Pebble game algorithms and sparse
graphs},  Discrete Mathematics.   308(8) (2008)  1425--37.


\bibitem{Lehman1964}
A. Lehman, {\it A solution to the Shannon switching game}, J. Soc.
Indust, Appl.  Math., 12 (1964) 687--725.


\bibitem{Lorea75}
 M. Lorea, {\it Hypergraphes et matroides}, Cahiers
Centrel Etud. Rech. Oper. 17 (1975) 289--291.


\bibitem{Lovasz}
L. Lov\'asz, Combinatorial Problems and Exercises,
North-Holland 1979. 


\bibitem{Lovasz70a}
 L. Lov\'asz, {\it A generalization of K{\H o}nig's
theorem,} Acta.  Math.  Acad.  Sci.  Hungar. 21 (1970) 443--446.


\bibitem{Lovasz76a}
L. Lov\'asz, {\it On two minimax theorems in graph
theory,} Journal of Combinatorial Theory (B) 21 (1976) 96--103.




\bibitem{Mader82}
W. Mader, {\it Konstruktion aller $n$-fach
kantenzusammenh\"angenden Digraphen}, Europ.  J. Combinatorics, 3
(1982) 63--67.

\bibitem{Menger}
K. Menger, {\it Zur allgemeinen Kurventheorie}, Fund. Math. 10 (1927) 96-–115.

\bibitem{MNW2007}
B. Mohar, R.J.  Nowakowski, and D.B.  West, {\it Research problems
from the 5th Slovenian Conference (Bled, 2003)}, Discrete Mathematics,
307(3-5) (2007) 650--658.






\bibitem{NWir}
C.St.J.A. Nash-Williams, {\it On orientations,
connectivity and odd vertex pairings in finite graphs}, Canad.  J.
Math. 12 (1960) 555--567.

\bibitem{Nash61}
C.St.J.A.  Nash-Williams, {\it Edge-disjoint spanning
trees of finite graphs}, The Journal of the London Mathematical
Society, 36 (1961) 445--450.

\bibitem{Nash64}
 C.St.J.A.  Nash-Williams, {\it Decomposition of
finite graphs into forests,} The Journal
of the London Mathematical Society, 39 (1964) 12.






\bibitem{Rado1962a}
 R. Rado, {\it A combinatorial theorem on vector spaces}, The Journal
of the London Mathematical Society, 37 (1962) 351--353.




\bibitem{Recskibook}
 A. Recski, Matroid Theory and its
Applications in Electric Network Theory and in Statics, Springer,
Berlin, 1989.

\bibitem{R-S}
I. S. Reed and G. Solomon, {\it Polynomial codes over certain finite fields}, Journal of the Society for Industrial and Applied Mathematics. 8 (2), (1960) 300-–304.




\bibitem{Schrijverbook} A. Schrijver, Combinatorial Optimization:
Polyhedra and Efficiency, Springer, 2003.  Vol 24. of the series
Algorithms and Combinatorics.







\bibitem{Tay1984}
T-S.  Tay, {\it Rigidity of multi-graphs.  I. Linking rigid bodies in
$n$-space},  Journal of Combinatorial Theory, Series B, 36(1)
(1984) 95--112.

\bibitem{Thomassen1989}
C. Thomassen, {\it Configurations in graphs of large minimum degree,
connectivity, or chromatic number}, Annals of the New York Academy of
Sciences. 555(1) (1989) 402--412.




\bibitem{Tutte61a} W.T.  Tutte, {\it On the problem of decomposing a
graph into $n$ connected factors}, The Journal
of the London Mathematical Society, 36 (1961)
 221--230.



\bibitem{Vidyasankar}
 K. Vidyasankar, {\it Covering the edge set of a
directed graph with trees}, Discrete Mathematics, 24 (1978) 79--85.




\bibitem{Whiteley96}
W. Whiteley, {\it Some matroids from discrete
applied geometry,} in:  Matroid Theory (J.E.  Bonin, J.G.  Oxley, and
B. Servatius, eds.)  Contemp.  Math., 197, Amer.  Math.  Soc.,
Providence, RI, (1996) 171--311.
\end{thebibliography}
\end{document}